\documentclass[aps,nofootinbib,preprintnumbers,citeautoscript]{revtex4}
\usepackage{amsmath,amssymb}
\usepackage{enumerate}
\usepackage{graphicx}
\graphicspath{{D:/researchdata/kapilpaper/arXiv/arxiv_Dynamics_of_entanglement_in_qubit_qutrit_with_x/figures/}}
\date{\today}

\begin{document}
\title{Dynamics of entanglement in qubit-qutrit with x component of DM interaction}

\author{Kapil K. Sharma$^\ast$ and  S. N. Pandey$^\dagger$  \\
\textit{Department of Physics, Motilal Nehru National Institute of Technology,\\ Allahabad 211004, India.} \\
E-mail: $^\ast$scienceglobal@gmail.com, $^\dagger$snp@mnnit.ac.in}

\begin{abstract}
In this present paper we study the entanglement dynamics in qubit A-qutrit B pair under x component of Dzyaloshinshkii-Moriya $(D_{x})$ by taking an auxiliary qubit C. Here we consider an entangled qubit-qutrit pair initially prepared in two parameter qubit-qutrit states and one auxiliary qubit prepared in pure state interacts with the qutrit of the pair through DM interaction. We trace away the auxiliary qubit and calculate the reduced dynamics in qubit A-qutrit B pair to study the influence of the state of auxiliary qubit C and $D_{x}$ on entanglement. We find that the state (probability amplitude) of auxiliary qubit does not influence the entanglement, only $D_{x}$ influences the same. The phenomenon of entanglement sudden death (ESD) induced by $D_{x}$ has also been observed. We also present the affected and unaffected two parameter qubit-qutrit states by $D_{x}$. 
\end{abstract}
\maketitle


\section{Introduction}
\label{intro}
Entanglement \cite{EPR1935,Neilsen2000,ent} is the most fascinating phenomenon in quantum information which has no classical counterpart. Many future quantum technologies are based on the same. However, various mathematical investigations are on the way to explore the inherited properties of entanglement. The phenomenon of entanglement is very much fragile with respect to environmental interactions. Quantum measurements also leads to decoherence which destroy the entanglement in quantum systems. If the entanglement is destroyed for finite interval of time then this phenomenon is called entanglement sudden death (ESD) \cite{YuEberly2004,YuEberly2006,YuEberly2009}. So the study of influence of interactions on entanglement in varieties of quantum systems is important for practical quantum information processing. Zhang et al. have studied the entanglement dynamics of qubit A-qubit B by taking a third qubit C which interact with the qubit $B$ under z component of DM interaction \cite{ZQiang1,ZQiang2,ZQiang3}. They have shown that DM interaction amplify and periodically kills the entanglement between two qubits (A and B) and by adjusting the state (probably amplitude) of third qubit $C$ and DM interaction strength, one can manipulate the entanglement and control the entanglement sudden death (ESD). They have also studied the same system by taking the x-component of the DM interaction $(D_{x})$ \cite{ZQiang4}. Here we mention that this kind of study is useful not only in qubit-qubit systems but it is also useful for hybrid quantum systems. Recently, we have studied the dynamics in hybrid qubit-qutrit systems by taking the DM interaction strength in z direction \cite{kk1,kk2,kk3}. DM interaction in different directions have different impact on  varieties of quantum states. So, it is interesting to study the dynamics of entanglement with x component of DM interaction.

Motivated, with the above studies we study the entanglement dynamics in hybrid qubit-qutrit systems with $D_{x}$. In the present paper we have considered a closed system of qubit A-qutrit B, which has been prepared in two parameter states, and an auxiliary qubit $C$, prepared in pure state, interacts with the qutrit $B$ of the closed system. The auxiliary quantum systems play an important role in quantum information processing. Many auxiliary quantum systems have been investigated which assist in improving the entanglement in quantum systems \cite{auxiliary_1,auxiliary_2,auxiliary_3,auxiliary_4}. We traced away the auxiliary qubit C to calculate the reduced system dynamics. Further, we study the influence of the state of auxiliary qubit C and $D_{x}$ on the entanglement between qubit A-qutrit B pair. We find that the state (probability amplitude) of third qubit $C$ do not influence the entanglement in two parameter qubit-qutrit states, only DM interaction influences the same. The $D_{x}$ is geometry dependent, so by changing the geometrical arrangement between interacting qutrit $B$ and qubit $C$, one can manipulate the entanglement between qubit-qutrit pair. Dzyaoshinshkii-Moriya interaction \cite{DMinteraction,DMmoriya,Tmoriya2_1960} is an anisotropic antisymmetric interaction investigated by taking into consideration the relativistic effects to describe the ferromagnetism of anti ferromagnetic crystals. Quantum spin chains are the important building blocks to execute the practical quantum information processing \cite{spin_chain}. Many spin chains with external magnetic filed and DM interaction have been investigated in varieties of configurations even at thermal conditions \cite{s1,s2}. Spin chains have also been studied by taking as a bath in the form of spin chains with DM interaction \cite{thermal1,thermal2,thermal3,thermal4,thermal5,thermal6,thermal7}. It has been observed that in some cases the DM interaction enhance the entanglement present in quantum spin chains. So DM interaction is a useful resource in quantum information processing. Recently we also have shown the efficacy of DM interaction to free the bound entangled states \cite{kk4,kk5} in quantum information processing. In the present study, we find that DM interaction does not kill the entanglement in some two parameter qubit-qutrit states, so the affected and unaffected states have also been reported here. The two parameter qubit-qutrit states  in $2\otimes n$ quantum systems have been investigated by Pyo Chi \cite{DongPyo} and its generalization in higher dimensions $(m\otimes n, n\geq m \geq 3)$ quantum systems by YaoMin \cite{DIYao}. Entanglement dynamics in two-parameter qubit–qutrit states under various channels has also been studied by Hai-Rui \cite{Hai-rui2012} and Hao and Fu \cite{Y_Hao}.

The plan of the paper is as follows. In Sect. 2, we present the Hamiltonian of the system considered. Two parameter qubit-qutrt  states its separable and non-separable regions and entanglement measure have been given in Sect 3. Sect 4 is devoted to the entanglement dynamics with the x component of DM interaction. Finally in Sect 5, we present our conclusion.   

\section{Hamiltonian of the system}
In this section, we present the Hamiltonian of the system. We consider the closed system of qubit $A$ and qutrit $B$ and, an auxiliary qubit $C$ which interact to the qutrit $B$ of the closed system via $D_{x}$. The schematic diagram of the system has been shown in Fig. 1. The Hamiltonian of the system can be written as
\begin{equation}
H=H_{AB}+H_{BC}^{int}, \label{eq:17}
\end{equation}
where $H_{AB}$ is the Hamiltonian of qubit $A$ and qutrit $B$ and $H_{BC}^{int}$ is the interaction Hamiltonian of qutrit (B) and qubit $C$. Here we consider uncoupled qubit A and qutrit B, so $H_{AB}$ is zero. Now the Hamiltonian becomes
\begin{eqnarray}
H=H_{BC}^{int}=\vec{D}.(\vec{\sigma_B} \times \vec{\sigma_C}),   \label{eq:18}
\end{eqnarray}
\begin{figure*}
        \centering
                \includegraphics[width=0.4\textwidth]{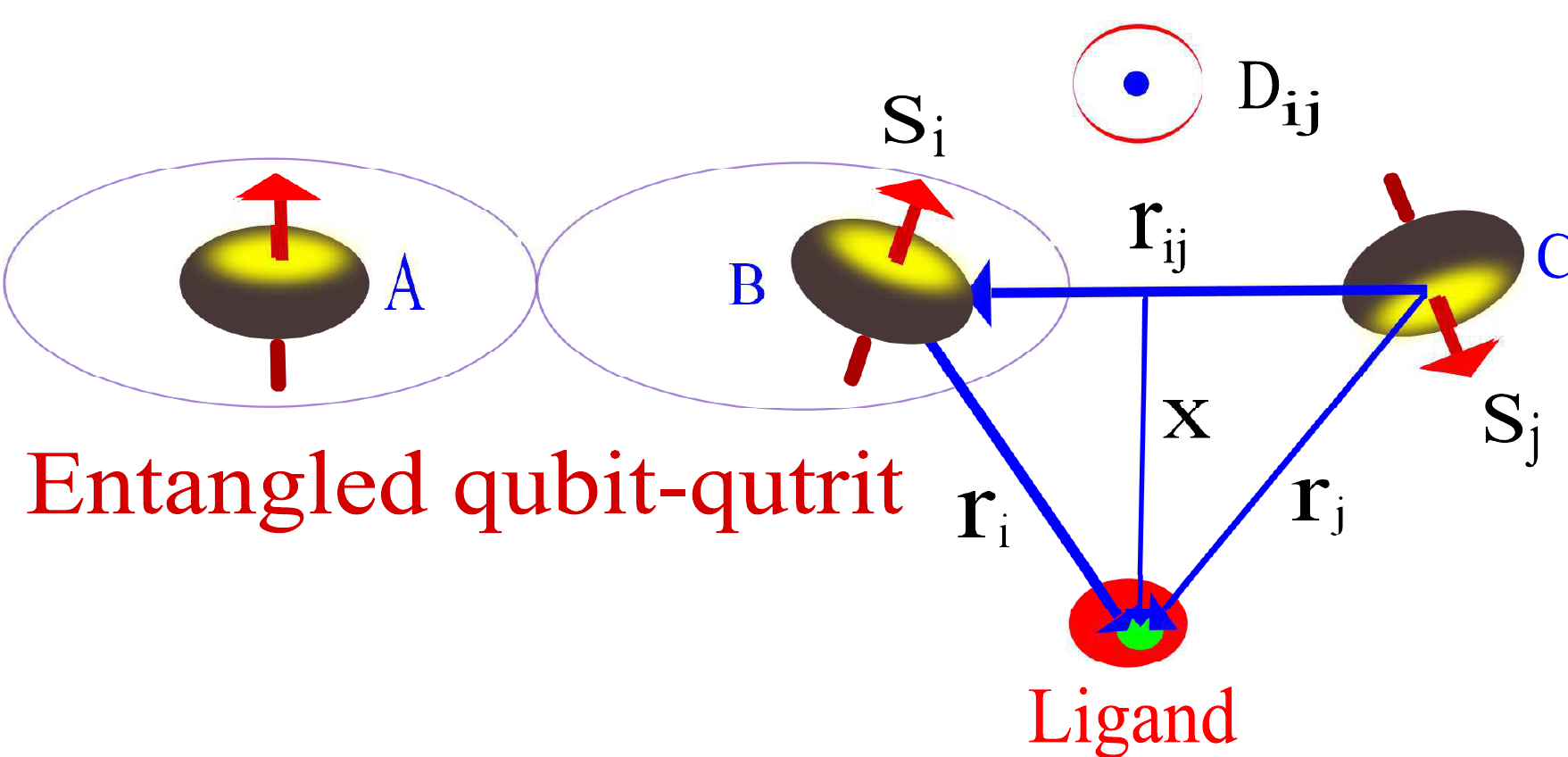}
                \caption{The system}
\end{figure*}
where $\vec{D}$ is DM interaction between qutrit $B$ and qubit $C$. Here $\vec{\sigma_B}$ is a vector associated with qutrit (B) whose components are Gell-Mann matrices and $\vec{\sigma_C}$ is the Pauli vector of qubit $C$ whose components are Pauli matrices. 
We assume that DM interaction exist along the x-direction only. So the Hamiltonian can be simplified as\\ 
\begin{equation}
H=D_{x}.(\sigma_B^Y \otimes \sigma_C^Z-\sigma_B^Z \otimes \sigma_C^Y),    \label{eq:19} \\ 
\end{equation}
where $\sigma_B^Y$ and $\sigma_B^Z$ are Gell-Mann matrices for qutrit $B$ and $\sigma_C^Y$ and $\sigma_C^Z$ are Pauli matrices of qubit $C$ respectively. The above Hamiltonian is a matrix having $6\times 6$ dimension and is easy to diagonalize by using the method of eigendecomposition. The unitary time evolution operator is easily commutable as 
\begin{eqnarray}
U(t)=e^{-i H t}, \label{eq:20}
\end{eqnarray}
which is also a $6 \times 6$ matrix. This matrix has been used to obtain the time evolution of density matrix of the system.
\section{Two parameter class of states and entanglement measure}
We describe the two parameter class of states of qubit-qutrit and reduced system dynamics in the following subsections 3.1 and 3.2 respectively.
\begin{figure*}
        \centering
                \includegraphics[width=0.4\textwidth]{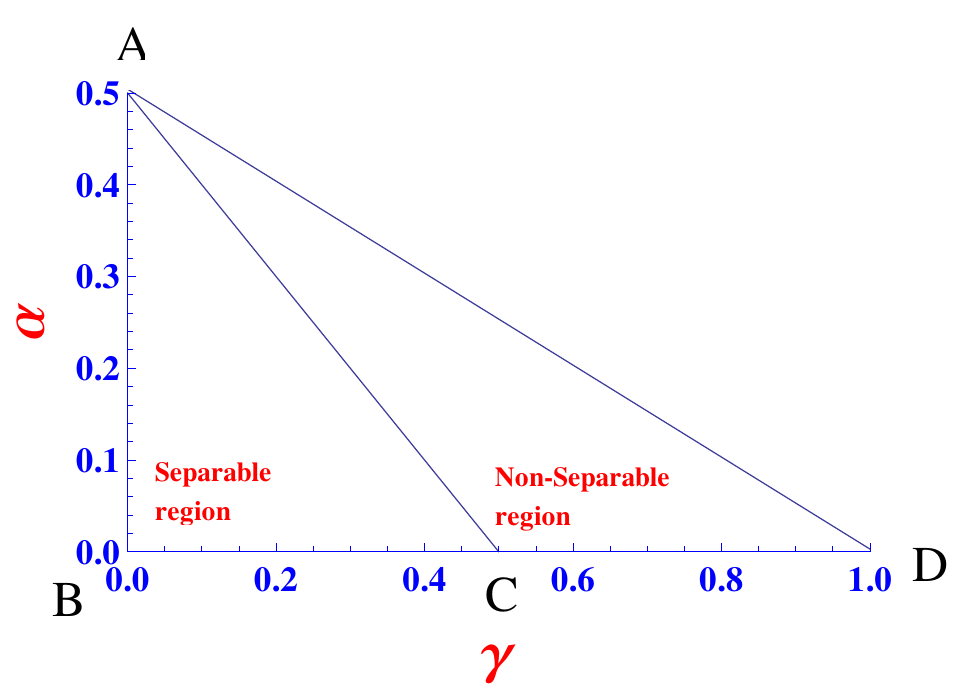}
                \caption{Separable and non-separable regions in two parameter qubit-qutrit states}
\end{figure*}
\subsection{Two parameter class of states}
The class of states in qubit-qutrit $(2\otimes 3)$ dimensional quantum system is given as \cite{DongPyo}
\begin{equation}
\widehat{\rho}_{AB}=\alpha(|02\rangle\langle 02|+|12\rangle\langle 12|)+\beta(|\phi^{+}\rangle\langle\phi^{+}|+|\phi^{-}\rangle\langle\phi^{-}|+|\psi^{+}\rangle\langle\psi^{+}|)+\gamma|\psi^{-}\rangle\langle\psi^{-}|,  \label{eq:55}
\end{equation}
where \\ 
\begin{equation}
|\phi^{\pm}\rangle=\frac{1}{\sqrt{2}}(|00\rangle\pm|11\rangle), \label{56} 
\end{equation}
\begin{equation}
|\psi^{\pm}\rangle=\frac{1}{\sqrt{2}}(|01\rangle\pm|10\rangle), \label{57} 
\end{equation} \\
and real parameters $\alpha,\beta,\gamma$ satisfy the following condition, 
\begin{equation}
2\alpha+3\beta+\gamma=1, \label{58}
\end{equation}
which is obtained by taking the trace of the density matrix given in Eq. (\ref{eq:55}). The negativity $N(\rho^{AB})$ of the two parameter class of states for $2\otimes 3$ dimension given in (5) can be obtained by using the formula
\begin{equation}
N(\rho^{AB})=\frac{1}{2}\left(\left\|\rho^{TA}\right\|{}_1-1\right), \label{eq:47}  \\
\end{equation} 
where $\rho^{TA}$ is the partial transpose of the reduced density matrix and $\parallel..\parallel_1$ denotes the trace norm.
The negativity is obtained as \\
\begin{equation}
N(\rho^{AB})=max\{0,2(\alpha+\gamma)-1\}. \label{59}
\end{equation} 
The two parameter class of state is separable within the limit $0\leq(\alpha+\gamma)\leq 1/2$ and non-separable within the limit $1/2<(\alpha+\gamma)\leq 1$. There are two triangular regions: separable and non-separable as depicted in Fig. 2. and the
separable region is bounded with the vertices $(\gamma=0, \alpha=0)$, $(\gamma=0, \alpha=1/2)$ and $(\gamma=1/2, \alpha=0)$. The non-separable region is bounded with the vertices $(\gamma=1/2, \alpha=0)$, $(\gamma=0, \alpha=1/2)$ and $(\gamma=1, \alpha=0)$. The vertices $(\gamma=0, \alpha=1/2)$ and $(\gamma=1/2, \alpha=0)$ are the common vertices of both seprable and non-separable regions. The line joining above mentioned vertices forms the boundaries of separable and non-separable regions. The separable region is bounded by the boundaries $BC$, $BA$ and $AC$, and non-separable region is bounded by the boundaries $CD$, $AC$ and $AD$. The boundary $AC$ is the common boundary of both the regions and states falling on this boundary belong to the separable region.
\section{Entanglement dynamics under the x-component of DM interaction $(D_{x})$}
To begin the study of entanglement dynamics we consider that the auxiliary qubit is prepared in pure state as below \\
\begin{equation}
\psi\rangle=c_{0}|0\rangle+c_{1}|1\rangle,
\end{equation}\\
with the normalization condition \\
\begin{equation}
c_{0}^{2}+c_{1}^{2}=1 \label{12}.
\end{equation}
To study the entanglement dynamics with the $D_{x}$ we need to calculate the reduce density matrix of the system. First we calculate the time evolution composite density matrix as $\rho(t)=U(t)\rho(0)U^{\dagger}(t)$, where $U(t)$ is the unitary time evolution operator given in Eq. (\ref{eq:20}) and $\rho(0)$ is the initial composite density matrix of qubit $A$, qutrit $B$ and qubit $C$, which can be written as $\rho(0)=\rho^{AB}\otimes \rho^{C}$. Here $\rho^{C}$ is the density matrix of the auxiliary qubit which is prepared in pure state. Further, we obtain the reduce density matrix by tracing away the auxiliary qubit $C$. The reduce density matrix is obtained for further calculations which is complicated to write here. We get the term $p=(c_{0}^{2}+c_{1}^{2})$ with every element of the matrix, so by imposing the normalization condition given in Eq. (\ref{12}), the factor disappear from the matrix and hence the probability amplitude of auxiliary qubit do not play any role in entanglement dynamics. Next we calculate the entanglement by using reduce density matrix with the help of negativity given in Eq. (9) along the boundaries $BC$, $BA$, $AC$, $CD$, $AD$ and the separable and non-separable regions. The study has been discussed in the subsequent sections.
\begin{figure*}
        \centering
                \includegraphics[width=1.0\textwidth]{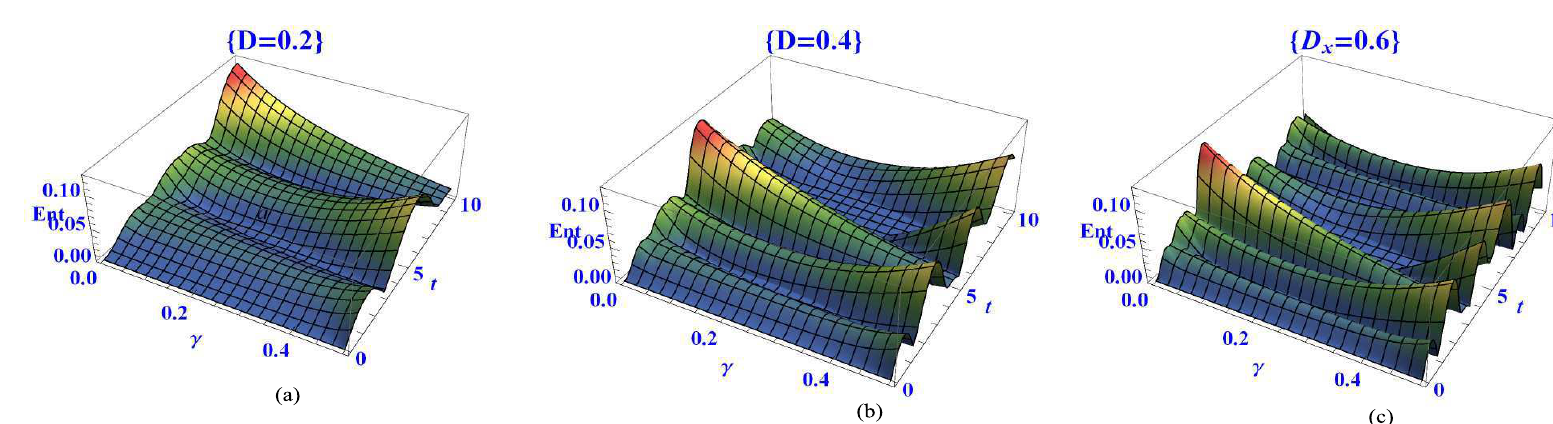}
                \caption{Entanglement plot along the boundary BC}
\end{figure*}
\begin{figure}
        \centering
                \includegraphics[width=1.0\textwidth]{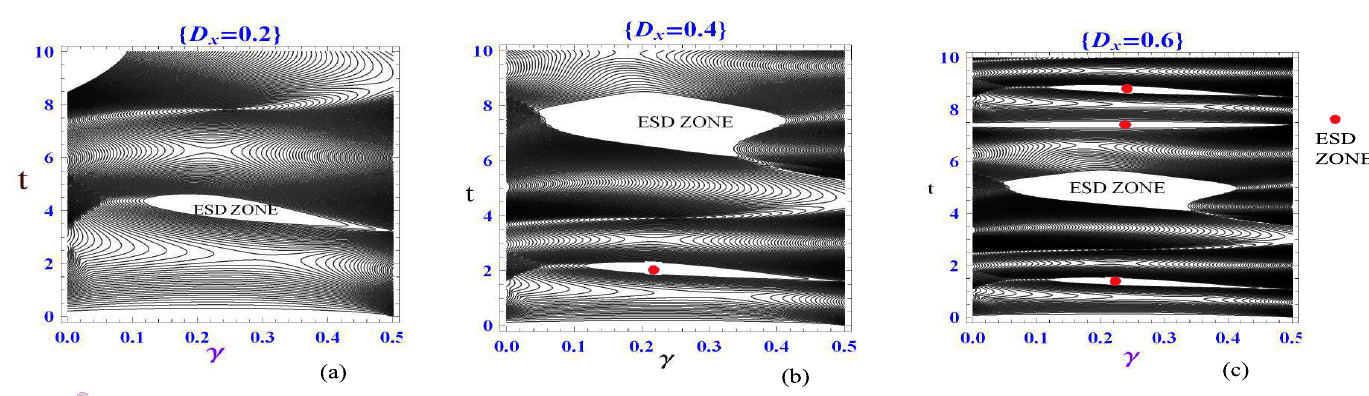}
                \caption{Contour plot along the boundary BC}
\end{figure}
\subsection{Entanglement dynamics along the boundary BC}
In this section, we present the study of entanglement dynamics along the boundary $BC$ which lies in separable region as dipicted in Fig 2. For $D_{x}=0$ there is no entanglement present in the quantum states lying along the boundary $BC$ as all the states are separable. We plot the entanglement for $D_{x}=0.2, 0.4$ and $0.6$ with associated contour plots in Fig. 3 and 4. First we plot the entanglement evolution for $D_{x}=0.2$ with $0\leq \gamma \leq 0.5$, which is shown in Fig. 3(a) and its corresponding contour plot is depicted in Fig. 4(a). The ESD is observed in contour plot as shown in Fig 4(a). The ESD zone lies within the range $0.11\leq \gamma \leq 0.50$ and within the time interval $3.44\leq t \leq 4.31$. Further, we plot the entanglement for $D_{x}=0.4$ in Fig. 3(b) and its corresponding contour plot in Fig. 4(b). By observing both the figures we conclude that the maximum amplitude of the entanglement remains same but the fluctuations of entanglement increases. Next we plot the results for $D_{x}=0.6$ in Fig. 3(c) and 4(c). From the contour plots we observe that increasing DM interaction strength produces more ESD zones and previous ESD zones shrinks. We conclude that increasing strength of DM interaction produces the entanglement in the states lying along the boundary $BC$ and maximum amplitude achieved by the entanglement is $0.1$. 
\begin{figure*}
        \centering
                \includegraphics[width=1.0\textwidth]{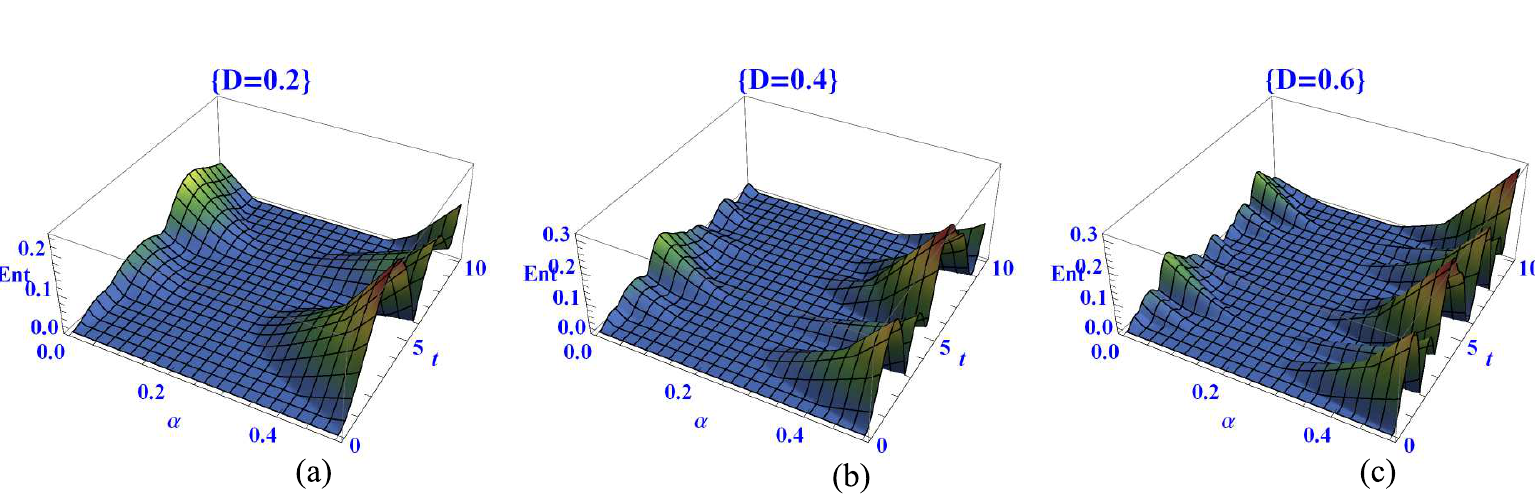}
                \caption{Entanglement plot along the boundary BA}
\end{figure*}
\begin{figure}
        \centering
                \includegraphics[width=1.0\textwidth]{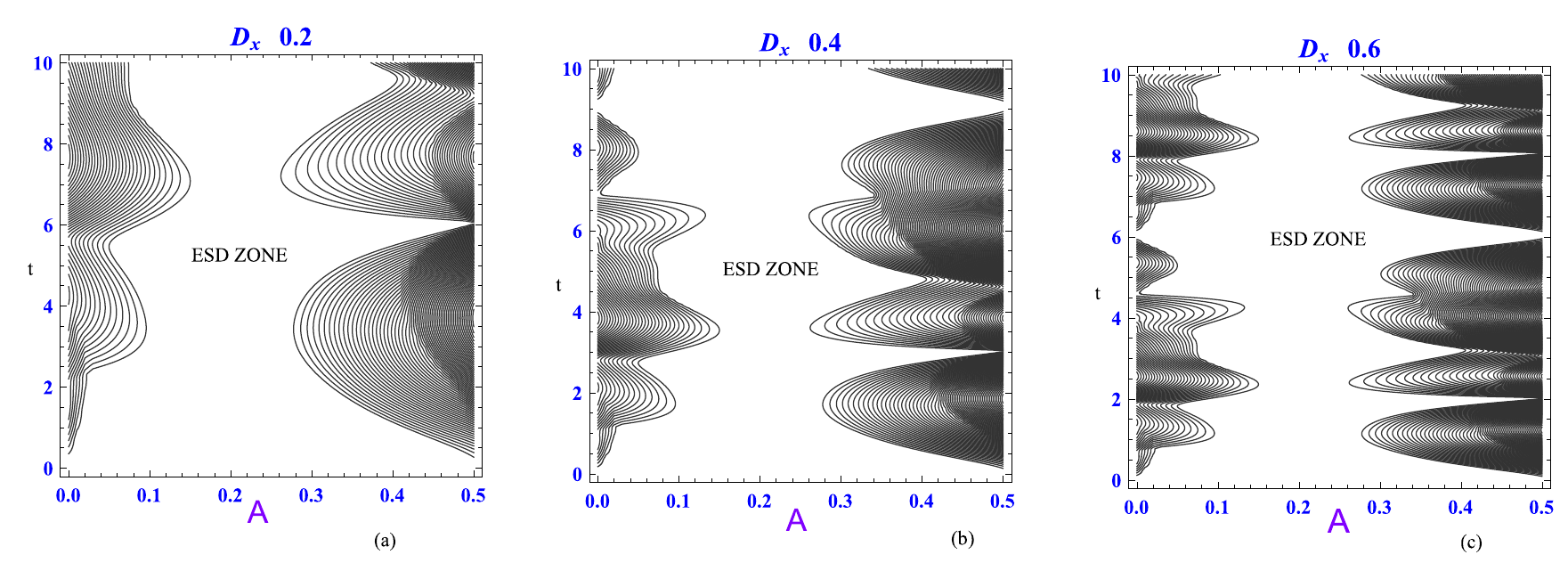}
                \caption{Contour plot along the boundary BA}
\end{figure}
\subsection{Entanglement dynamics along the boundary BA}
The entanglement dynamics along the boundary $BC$ has been discussed in the previous section. In this section, we study the entanglement dynamics through DM interaction along the boundary $BA$. All the states lying along this boundary are separable. We plot the entanglement evolution lying along this boundary with $0\leq \alpha \leq 0.5$. The results are plotted in Figs. 5 and 6. First we plot the result for $D_{x}=0.2$ in Fig. 5(a) and its corresponding contour plot in Fig. 6(a). We observe that for initial values of $\alpha$ the entanglement is produced  but as the value of $\alpha$ increases the entanglement dies out with the advancement of time. We observe the ESD zone in contour plot as shown in Fig. 6(a). The maximum amplitude of entanglement achieved is $0.2$. For $D_{x}=0.4$ we show the evolution of entanglement in Fig. 5(b) and its corresponding contour plot in Fig. 6(b). Here we observe that the maximum value of entanglement is $0.3$. The ESD is also seen in Fig. 6(b). Next we plot the entanglement with $D_{x}=0.6$ in Fig. 5(c) and its corresponding contour plot in Fig. 6(c). We observe that the maximum amplitude of entanglement is now stabilized with the increasing value of DM interaction strength. In the corresponding contour plot we again observe ESD zones. 
\begin{figure*}
        \centering
                \includegraphics[width=1.0\textwidth]{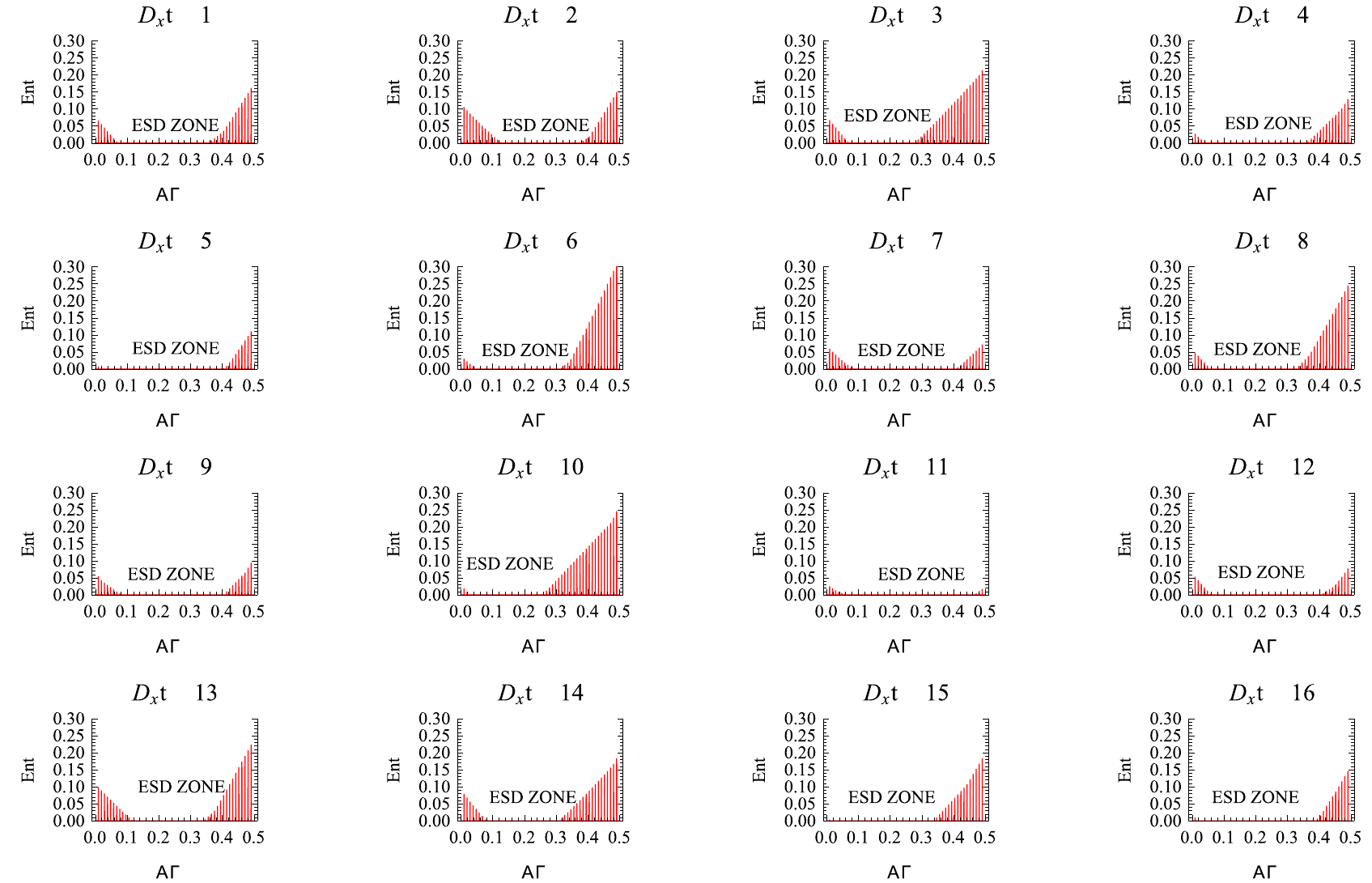}
                \caption{Entanglement plot inside the region ABC}
\end{figure*}
\subsection{Entanglement dynamics inside the region ABC including the boundary AC}
In this section, we discuss the entanglement evolution inside the region $ABC$ including the boundary $AC$. The region satisfy the condition $0\leq (\alpha+\gamma) \leq 0.5$. It is very complicated to obtain the results by varying the parameters $\alpha$, $\gamma$, $D_{x}$ and $t$ all together so we plot the entanglement evolution for different values of the parameter $D_{x}t$. The results are shown in Fig. 7. We observe that with the increasing value of $D_{x}t$, the ESD zones periodically shrink and expand. By observing all the figures we find that the maximum amplitude of entanglement is $0.3$ (which is obtained for $D_{x}t=0.6$) and entanglement periodically oscillates on both the sides of the middle point $(\alpha+\gamma)=0.25$. Further, we find that the maximum ESD zone is produced with $D_{x}t=11$ with the amplitude $0.02$ of entanglement. This result is shown in Fig. 8(a).
\subsection{Entanglement dynamics inside the region $ACD$} 
In this section, we describe the results obtained for entanglement evolution for the region $ACD$. The result has been shown in Fig. 8(b). We plot the entanglement evolution for different values of $D_{x}t$ and observe that as the value of $D_{x}t$ increases the amplitude of entanglement increases. And it is important to note that there is no ESD found in this region $ACD$. We here recall the result obtained with the z component of DM interaction of earlier study \cite{kk2}. In both the cases we observe that the states lying in the region $ACD$ do not suffer from ESD.
\begin{figure}
        \centering
                \includegraphics[width=1.0\textwidth]{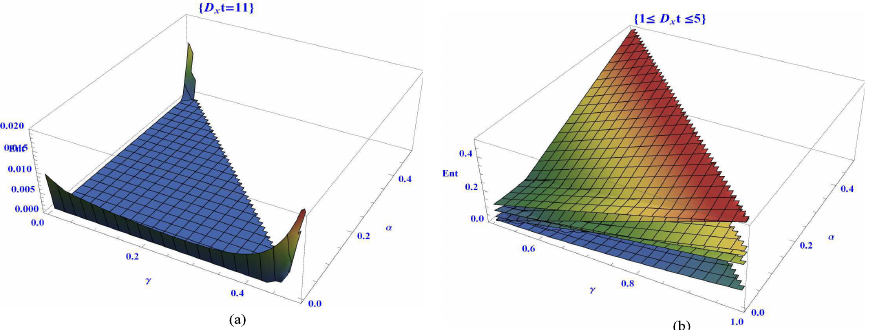}
                \caption{(a) Maximum entanglement plot with $D_{x}t=11$ (b) Entanglement plot inside the region ACD}
\end{figure}
\section{Conclusion}
In this paper, we present the study of entanglement dynamics in two parameter qubit-qutrit states with the x-component of DM interaction by taking an auxiliary qubit. We have studied the entanglement dynamics in separable and non-separabe regions and along thier associated boundaries. First we observe that as the value of DM interaction strength increases the entanglement is produced in the separable region with ESD. The maximum amount of entanglement achieved in this region is $0.3$. Next we present the dynamics along the boundary $BC$ of the separable region. We observed that the states lying along this boundary goes under ESD as time advances. As the value of DM interaction strength increases, the oscillations of entanglement increases and the ESD zones periodically shrink and expand. The maximum entanglement achieved along the boundary $BC$ is $0.1$. Further, we study the entanglement dynamics along the boundary $BA$ and found that the states lying along this boundary incorporate ESD zones with periodically shrinking and expanding effects. The maximum entanglement produced along this boundary is $0.3$. Next we study the entanglement dynamics inside the region $ABC$ including the boundary $AC$ with different values of the parameter $D_{x}t$. As the value of the parameter $D_{x}t$ increases the ESD zone periodically shrink and expand inside this region also. We observed maximum ESD zone inside this region with $0.05 \leq (\alpha+\gamma)\leq 0.45$ at the parameter value $D_{x}t=11$. We also present the entanglement dynamics inside the non-separable region and plot it for different values of the parameter $D_{x}t$. We find that the increasing strength of DM interaction increases the amplitude of the entanglement present in this region but do not produce the ESD zones. 

We get maximum amount of entanglement as 0.15 due to z component of DM interaction, x component of DM interaction is producing maximum entanglement as 0.30 in separable region ABC. By comparing the maximum entanglement produced by x and z components of DM interaction in separable region ABC, we conclude that x component of DM interaction is producing more amount of maximum entanglement. However in the region ACD, the x and z components are producing same amount of entanglement.   

\end{document}